\documentclass[pra,twocolumn,floatfix,showpacs,superscriptaddress]{revtex4}
\usepackage{amsmath}
\usepackage{amsfonts}
\usepackage{graphicx}

\begin{document}

\title{Entanglement on macroscopic scales 
in a resonantly laser-excited atomic ensemble}
\author{S. Camalet}
\affiliation{Laboratoire de Physique Th\'eorique 
de la Mati\`ere Condens\'ee, UMR 7600, Sorbonne 
Universit\'es, UPMC Univ Paris 06, F-75005, 
Paris, France}
\begin{abstract}
We show that two groups of slow two-level atoms 
in a weak resonant laser field, are entangled. 
The considered groups can be separated by 
a macroscopic distance, and be parts of a larger 
atomic ensemble. In a dilute regime, for two very 
distant groups of atoms, in a plane wave laser beam, 
we determine the maximum attainable entanglement 
negativity, and a laser intensity below which they are 
certainly entangled. They both decrease with increasing 
distance between the two groups, but increase 
with enlarging groups sizes. As a consequence, 
for given laser intensity, far separated groups of atoms 
are necessarily entangled if they are big enough.
\end{abstract}


\maketitle 

\section{Introduction}
  
The impact of the environment on quantum entanglement 
is manifold. On the one hand, an initial entanglement 
between independent systems, can be destroyed by 
their coupling to their surroundings. This fragility of quantum 
entanglement is substantiated by the fact that it can vanish 
in finite time, as first shown for two-level atoms in distinct 
vacuum cavities \cite{YE}. On the other hand, the environment 
mediates interactions, Hamiltonian or not, between 
the considered systems, and can then induce correlations 
between them, and hence potentially entanglement. 
It has been shown that finite entanglement can develop 
between two initially uncorrelated two-level systems, 
or qubits, sharing the same surroundings, but otherwise 
uncoupled \cite{Br,J,BFP,FT,MCNBF}. However, in 
the realistic case of a finite separation between the two 
systems, this effect is only transient if the environment is 
in thermal equilibrium \cite{FT,MCNBF}. 
 
This is not the case when the surroundings does not reduce 
to a thermal bath. The systems steady state, reached 
asymptotically from any initial state, can present finite 
entanglement, as has been shown for two qubits in diverse 
environments, such as the electromagnetic vacuum and 
a resonant laser field \cite{CKS,ATF}, two heat reservoirs 
at different temperatures \cite{EPJB}, and the electromagnetic 
field emitted by two bodies at different temperatures 
\cite{BA1,BA2}. The entanglement obtained in these works, 
can be essentially traced back to one of the two familiar 
features of a pair of infinitely close qubits, which are, 
the decoupling, from the environment, of the so-called 
subradiant state, and the divergent energy shifts 
of the single-excitation levels \cite{CDG2,D,L2}. 
Experimentally, transient entanglement of two atoms 
has been generated using the Rydberg blockade 
mechanism, which relies on similar dipole-dipole energy 
shifts, but of double-excitation levels \cite{Wea,Zea}. 
Another approach consists in starting with multi-level 
atoms coupled to one or several monochromatic fields, 
and possibly to static fields or to a common cavity mode,
and then reducing them to two-level systems by adiabatic 
elimination of upper levels in an appropriate 
regime \cite{PSC,MPC,RKS}. Following the proposal 
of Ref.\cite{MPC}, long-lived entanglement of two atomic 
ensembles separated by 50 cm, has been observed 
experimentally \cite{KM,MK}. 

\begin{figure}
\centering \includegraphics[width=0.45\textwidth]{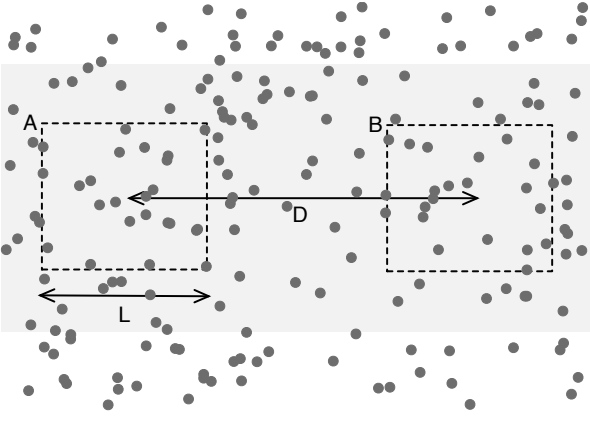}
\caption{\label{fig:at} Schematic representation of two subgroups 
of atoms, A and B, of characteristic size $L$, and separated 
by a distance $D$, of a larger atomic ensemble, partially 
illuminated by a resonant laser beam. A and B are entangled, 
for $D \simeq 1$ m and $L \simeq 50 \mu$m, for example, 
see Sec.~\ref{sec:Lre}.}
\end{figure}

In this paper, we are concerned with the entanglement 
generated by a resonant laser field in an atomic ensemble. 
Entangled steady states of systems of many driven qubits 
have been obtained for highly simplified interaction models, 
in which each qubit is coupled identically to every other 
one \cite{MEK}, or is coupled only to its nearest neighbours 
in a chain \cite{EPJB2}. We consider here two-level atoms 
evolving under the influence of a laser field and 
of the obviously present electromagnetic vacuum. 
Lehmberg's master equation is used to describe 
the dynamics of the atoms internal state 
\cite{FT,CKS,ATF,L2,MPC,L1}. For two atoms, the resulting 
steady state is separable or entangled, 
depending on the relative strength of the laser amplitude 
and vacuum-mediated interaction \cite{CKS,ATF,MEK}. 
We focus on the regime of low laser intensities, in which 
entanglement is long-range. As we will see, the underlying 
physical origin of the found entanglement, 
is that, for weak laser fields, the atoms internal dynamics 
is dominated by the dipole-dipole interaction, 
laser photon absorption and collective radiative decay, 
which all preserve the state purity. 
As a result, the atoms steady state is practically pure, 
and correlated, and hence entangled. As shown in 
the following, this remains true for large groups of atoms, 
and even if they are surrounded by other identical atoms, 
as illustrated in Fig.\ref{fig:at}.

The rest of the paper is organized as follows. 
The Hamiltonian used to describe laser-excited two-level 
atoms, and the approximations leading to Lehmberg's 
master equation, are presented in the next section. 
In Sec.~\ref{sec:Issa}, the steady internal state 
of slow-moving atoms in a weak resonant laser field, 
is determined, and two of its features, which are 
of particular importance for entanglement, 
are discussed. In Sec~\ref{sec:E}, 
the entanglement of any two subgroups of atoms is 
studied. It is shown that there is a laser intensity 
threshold, which depends on the considered atoms, 
below which the two subgroups are entangled. 
More quantitative results are derived, in a dilute regime, 
for macroscopically distant groups of atoms. Finally, 
in the last section, we summarize our results, and mention 
some questions raised by our study.

\section{Master equation for laser-excited two-level atoms} 
\label{sec:Meletla}

We consider an ensemble of two-level atoms evolving 
under the influence of a laser field. Within the dipolar 
approximation for the coupling to 
the electromagnetic field \cite{CDG2}, and a semiclassical 
approximation for the motion of the atoms \cite{DC}, 
the dynamics of the atoms internal state is governed 
by the Hamiltonian
\begin{eqnarray}
H &=& H_{e} + \omega_0 
\sum_{\mu} \sigma^\dag_\mu 
\sigma^{\phantom{\dag}}_\mu \label{H} \\
&~&-\sum_{\mu}
(\sigma^{\phantom{\dag}}_\mu+\sigma^\dag_\mu) 
\Big[ {\bf d}  \cdot {\bf E} ({\bf r}_{\mu}) 
+\frac{\Omega}{2} \mathrm{Re} \big(w({\bf r}_\mu) 
e^{-i  \omega t} \big) \Big] , \nonumber
\end{eqnarray} 
where $\omega_0$ is the atomic resonance 
frequency, ${\bf r}_\mu$ is the classical position 
of atom $\mu$, and $\Omega$ is proportional 
to the laser field amplitude. Throughout this paper, 
units are used in which $\hbar=1$. 
The Hamiltonian $H_e$ and field ${\bf E}$ read, 
respectively, $H_e=c \int \mathrm{d^3 k} |{\bf k}| 
(a^{\dag}_{{\bf k}1} a^{\phantom{\dag}}_{{\bf k}1} 
+a^{\dag}_{{\bf k}2} a^{\phantom{\dag}}_{{\bf k}2})$, 
and 
\begin{equation}
{\bf E} ({\bf r}) = \int \mathrm{d^3 k} 
\left(\frac{-c|{\bf k}|}{16 \pi^3 \epsilon_0}\right)^{1/2} 
\sum_{p=1,2} {\bf e}_{{\bf k} p} 
e^{i{\bf k} \cdot {\bf r}} a^{\phantom{\dag}}_{{\bf k}p}
+ h.c. , \label{E}
\end{equation} 
where $c$ is the speed of light, $\epsilon_0$ 
is the vacuum dielectric permittivity, 
${\bf e}_{{\bf k} 1}$ and ${\bf e}_{{\bf k} 2}$ 
are unit vectors orthogonal to ${\bf k}$ and 
to each other, and the electromagnetic field 
operators $a_{{\bf k} p}$ satisfy the bosonic 
commutation relations 
$[a^{\phantom{\dag}}_{{\bf k} p},a^{\dag}_{{\bf k}' p'}]
=\delta_{p p'} \delta({\bf k}-{\bf k}')$. 
The atomic operator $\sigma_\mu$ is defined by 
$\sigma_\mu=|g\rangle_\mu {_\mu\langle e|}$ 
where $|g\rangle_\mu$ and $|e\rangle_\mu$ 
are, respectively, the ground and excited states of 
atom $\mu$. The vector ${\bf d}=
{_\mu\langle e|} {\bf D}_\mu |g\rangle_\mu$ 
where ${\bf D}_\mu$ is the dipole moment of atom 
$\mu$, is assumed real and the same for all the atoms. 
The spatial function $w$ depends on the laser beam 
considered, $w({\bf r})=\exp(i{\bf K}\cdot{\bf r})$ 
for a plane wave of wave vector ${\bf K}$, for example. 

For fixed positions ${\bf r}_\mu$, the timescales relevant 
to the dynamics of the atoms internal state $\rho$ 
are $\omega_0^{-1}$, $|{\bf r}_\mu- {\bf r}_\nu|/c$, 
$\Omega^{-1}$, and $\Gamma^{-1}$ where 
$\Gamma = |{\bf d}|^2\omega_0^3 / 3\pi\epsilon_0c^3$ 
is the spontaneous decay rate of an isolated atom 
\cite{L1,S,TGCB}. In the following, we consider laser 
intensities such that $\Omega \ll \Gamma$. 
The ratio $\Gamma/\omega_0$ is of the order 
of $\alpha^3$ where $\alpha \simeq 7.10^{-3}$ is 
the fine-structure constant \cite{CDG2}. Thus, for 
distances $|{\bf r}_\mu- {\bf r}_\nu| \ll k_0^{-1}\alpha^{-3}$ 
where $k_0=\omega_0/c$, the timescale $\Gamma^{-1}$ 
and $\Omega^{-1}$ are very long compared to the other 
ones, and it can be shown that the time evolution of $\rho$ 
is well described by the master equation 
\begin{equation}
\partial_t \rho = -\Gamma {\cal L}(\{ {\bf r}_\mu \}) \rho 
-i [H_a+H_l,\rho] , \label{meqg}
\end{equation}
where $H_a$ and $H_l$ are, respectively, the second 
and last terms of Hamiltonian \eqref{H}. 
The superoperator ${\cal L}$ is defined by 
\begin{equation}
{\cal L} \varrho =  \sum_{\mu , \nu} \left[ z_{\mu \nu}
\sigma^\dag_\mu\sigma^{\phantom{\dag}}_\nu \varrho 
+  z_{\mu \nu}^*
\varrho \sigma^\dag_\mu\sigma^{\phantom{\dag}}_\nu
-2\gamma_{\mu \nu} 
\sigma^{\phantom{\dag}}_\mu \varrho \sigma^\dag_\nu 
\right] , \label{L}
\end{equation} 
where $\varrho$ is any matrix, 
$\gamma_{\mu \nu} =\mathrm{Re}z_{\mu \nu}$, 
$z_{\mu \mu}=1/2$, and, for $\mu \ne \nu$, 
\begin{equation}
z_{\mu \nu} = \frac{3}{4}\frac{e^{i r}}{r^3}
\Big\{ \big[1-3 ({\bf \hat d} \cdot {\bf \hat r})^2 \big](i+r)
-i\big[1-({\bf \hat d} \cdot {\bf \hat r})^2\big] r^2 \Big\} , 
\label{gamome}
\end{equation}
with $r=k_0 |{\bf r}_\mu- {\bf r}_\nu|$, 
${\bf \hat r}=({\bf r}_\mu- {\bf r}_\nu)/|{\bf r}_\mu- {\bf r}_\nu|$, 
and ${\bf \hat d}={\bf d}/|{\bf d}|$ \cite{S,L1}. For atoms 
moving with velocities $\partial_t {\bf r}_\mu 
\ll c/k_0 |{\bf r}_\nu- {\bf r}_\xi|$, equation \eqref{meqg} 
remains valid with the time-dependent positions 
${\bf r}_\mu(t)$ \cite{TGCB}. Thus, for typical values of 
$\omega_0$ and $\Gamma$, and velocities of the order 
of 10 m.s$^{-1}$, equation \eqref{meqg} is relevant 
for interatomic distances as large as some decimeters.

\section{Internal state of slow atoms 
in a weak resonant laser field} \label{sec:Issa}

We consider atoms velocities such that 
\begin{equation}
\partial_t {\bf r}_\mu \ll \Gamma k_0^{-1} . \label{vel}
\end{equation}
In terms of the temperature $T$ of the atomic ensemble,
this condition can be rewritten as $T/A \ll 1$ K where 
$A$ is the mass number of the atoms. Since 
the characteristic length scale of both 
$H_l$ and ${\cal L}$, is $k_0^{-1}$, the displacements 
of the atoms during a time interval of length $\Gamma^{-1}$, 
can be neglected in equation \eqref{meqg}. Consequently, 
at each instant $t$, $\rho$ is essentially equal to 
the asymptotic solution of this equation with atoms positions 
${\bf r}_\mu (t)$ assumed fixed. It is hence of the form 
$\rho=\sum_p{\rho}_p \exp(-ip \omega t)$. 
Due to the small value of $\Gamma/\omega_0$, the matrices 
$\rho_p$ are practically given by their zeroth order expansions 
in this ratio. Thus, they obey 
$[H_a,\rho_p]=p\omega_0\rho_p$, and are determined by 
\begin{equation}
i{\cal L} \rho_p + p\delta  \rho_p +  \eta \big[W,\rho_{p+1}]
+ \eta [W^\dag ,\rho_{p-1} \big] = 0 ,  \label{eqsp}
\end{equation}
where $\delta = (\omega-\omega_0)/\Gamma$ is 
a dimensionless laser detuning, which is assumed to be 
at most of order unity, $\eta=\Omega/2\Gamma$ 
and $W = \sum_\mu \sigma_\mu w({\bf r}_\mu)^*$. 
Note that only the rotating wave part of $H_l$, i.e., 
$\Omega W \exp(i\omega t)/2+h.c.$, appears in 
these equations.
 
\subsection{Low laser intensity perturbative solution} 
 
As we are concerned with laser intensities such that 
$\Omega \ll \Gamma$, we solve equations \eqref{eqsp} 
perturbatively in the ratio $\eta$. To do so, we expand 
the Fourier components $\rho_p$ as 
$\rho_p=\rho_p^{(0)}+\eta \rho_p^{(1)}+ \ldots$, 
where 
$\rho_p^{(0)}=\delta_{p0} | {\cal G} \rangle \langle {\cal G} |$, 
 with $|{\cal G} \rangle = \otimes_\mu | g \rangle_\mu$, 
 is the solution to eq.\eqref{eqsp} for $\eta=0$. 
From eq.\eqref{eqsp}, the successive $\rho_p^{(n)}$ are 
related by
\begin{equation}
(i{\cal L}+ p\delta) \rho_p^{(n+1)}  =- \big[W,\rho^{(n)}_{p+1}]
- [W^\dag ,\rho^{(n)}_{p-1} \big] .  \label{rr}
\end{equation}
Note that, strictly speaking, the regime of validity 
of the following results, is $\alpha^3  \ll \eta \ll 1$, since, 
in the derivation of eq.\eqref{eqsp}, terms of 
the order of $(\Gamma/\omega_0)^2$ have been 
neglected, whereas terms of the order 
of $\Omega/\omega_0$ have been kept. 

Using the recursive relation \eqref{rr}, 
we find, up to second order in $\eta$, 
\begin{equation}
\rho=\big(\langle \psi | \psi \rangle^{-1} 
| \psi \rangle\langle \psi | \big)^{[2]} ,
\label{gasstate}
\end{equation} 
where the superscript $^{[2]}$ means that only 
terms up to second order are kept, and
\begin{equation}
| \psi \rangle = |{\cal G} \rangle 
+ \eta \sum_\mu u_\mu |\mu \rangle 
+ \eta^2 \sum_{\mu < \nu} 
(u_\mu u_\nu + v_{\mu \nu}) |\mu \nu \rangle ,
\label{psi}
\end{equation}
with $|\mu \rangle = \sigma^\dag_\mu |{\cal G} \rangle$ and 
$|\mu \nu \rangle = \sigma^\dag_\mu |\nu \rangle$, 
see Appendix~\ref{app:Pse7}. 
The components $u_\mu$ and $v_{\mu \nu}$ obey
\begin{eqnarray}
 \sum_\xi z_{\mu \xi} u_\xi -i \delta u_\mu 
 &=& i w_\mu ,  \label{eqcoeff}  \\
\sum_\xi \big( z_{\mu \xi} {\tilde v}_{\xi \nu} 
+ z_{\nu \xi} {\tilde v}_{\xi \mu} \big) -2i\delta v_{\mu \nu} 
&=& z_{\mu \nu} \big(u_\mu^2+u_\nu^2 \big) , \nonumber
\end{eqnarray}
where $\mu < \nu$, $w_\mu=w({\bf r}_\mu)\exp(-i\omega t)$, 
${\tilde v}_{\mu \nu}$ equals $0$ for $\mu=\nu$, 
$v_{\mu \nu}$ for $\mu<\nu$, and $v_{\nu \mu}$ 
for $\mu>\nu$. 

The state of atom $\mu$, which reads 
$\rho_\mu=| \phi \rangle_\mu {_\mu\langle} \phi |
-\eta^2 |u_\mu|^2 |g \rangle_\mu {_\mu\langle} g |$ with 
$| \phi \rangle_\mu=|g \rangle_\mu+\eta u_\mu |e \rangle_\mu$, 
is determined by $u_\mu$, and the correlations between atoms 
$\mu$ and $\nu$ are determined by $v_{\mu \nu}$, since
$\rho_{\mu \nu} - \rho_{\mu} \otimes \rho_{\nu} =
\eta^2 v_{\mu \nu}^* \sigma_\mu \sigma_\nu +  h.c.$ 
where $\rho_{\mu \nu}$ is the state of the pair of atoms 
$\mu$ and $\nu$. Note that these correlations vanish 
if the mutual influence between the atoms, mediated 
by the electromagnetic vacuum, is neglected, 
since equations \eqref{eqcoeff} give $v_{\mu \nu}=0$ 
for $z_{\mu \nu}=\delta_{\mu\nu}/2$. 
In the absence of vacuum-mediated interaction, entanglement 
between atoms can still be generated using photons, but not 
with a simple laser field \cite{G,vE,CDLK}. 

\subsection{Schr\"odinger-like equation}

The fact that $\rho$ coincides, up to second order, with a pure 
state, plays an essential role in the following. The origin of 
this effective purity can be understood as follows. The first 
two terms of the superoperator \eqref{L}, which describe 
the dipole-dipole interaction between the atoms and 
the decay of the excited atomic levels due 
to spontaneous emission, can be interpreted in terms 
of an effective complex Hamiltonian. This is not the case 
of the last one, which accounts for the populating, 
by spontaneous emission, of 
$\langle k | \rho | l \rangle$ where $| k \rangle$ 
and $| l \rangle$ are eigenstates of $H_a$, from 
matrix elements $\langle k' | \rho | l' \rangle$ such that 
$\epsilon_{k'}>\epsilon_k$ where 
$\epsilon_k=\langle k | H_a | k \rangle$. 
However, for small $\eta$, and in the long time regime, 
this process essentially does not contribute to $\rho$, 
since the order, in $\eta$, of $\langle k | \rho | l \rangle$ 
increases with $\epsilon_k$. Up to second order, 
it only results in  a correction to the ground state 
population  $\langle {\cal G} | \rho | {\cal G} \rangle$, 
which simply ensures the normalisation of $\rho$. 
Due to the decline of $\langle k | \rho | l \rangle$ with 
increasing $\epsilon_k$, 
stimulated emission is also negligible. Consequently, 
equations \eqref{eqsp} can be approximated by 
the Schr\"odinger-like equation
\begin{equation}
\partial_t | \psi \rangle = -\Big[ i H_a 
+ \Gamma \sum_{\mu , \nu}  z_{\mu \nu}
\sigma^\dag_\mu\sigma^{\phantom{\dag}}_\nu
-i\frac{\Omega}{2} e^{-i\omega t}W^\dag \Big] 
| \psi \rangle , \label{effSeq}
\end{equation}
where the last term describes laser photon absorption. 
This equation is satisfied, up to second order, by a state 
of the form \eqref{psi}, provided $u_\mu$ and 
$v_{\mu \nu}$ fulfill eq.\eqref{eqcoeff}.

\subsection{State of a subensemble} 

An important property of the state \eqref{gasstate} is that 
the ensuing state of any subensemble of atoms, 
is given by an expression of the same form. 
Consider the system S consisting 
of the atoms $\mu =1, \ldots, n$, and the complementary 
system ${\overline {\mathrm{S}}}$ consisting of all 
the other atoms. The pure state \eqref{psi} 
can be expanded on the basis 
$\{ |{\cal G} \rangle_{\overline {\mathrm{S}}} , 
\sigma_\mu^\dag |{\cal G} \rangle_{\overline {\mathrm{S}}}, 
\ldots \}$ of system ${\overline {\mathrm{S}}}$, 
where $|{\cal G} \rangle_{\overline {\mathrm{S}}}
= \otimes_{\mu > n} | g \rangle_\mu$, 
as $| \psi \rangle = |{\cal G} \rangle_{\overline {\mathrm{S}}} 
| \psi \rangle_\mathrm{S} + \ldots$. The state 
$| \psi \rangle_\mathrm{S}$ is given by expression \eqref{psi}, 
but with sums running only over the first $n$ atoms, and 
$|{\cal G} \rangle$, $|\mu \rangle$ and $|\mu \nu \rangle$, 
replaced by the corresponding states for system S. The point 
is that the following terms in the above expansion 
of $| \psi \rangle$, either do not contribute to 
the second-order state $\rho_\mathrm{S}$ of system S, 
or contribute only 
to a correction to the population 
${_\mathrm{S}\langle} {\cal G} | \rho_\mathrm{S} 
| {\cal G} {\rangle_\mathrm{S}}$, 
which simply ensures the normalisation of 
$\rho_\mathrm{S}$. Consequently, $\rho_\mathrm{S}$ 
is given by eq.\eqref{gasstate} with $| \psi \rangle$ 
replaced by $| \psi \rangle_\mathrm{S}$.

\section{Entanglement} \label{sec:E}

In this section, we discuss the entanglement of any two 
subgroups of atoms, say A and B, such as 
those schematically depicted in Fig.\ref{fig:at}. 
A sufficient, but in general not necessary, 
condition for A and B to be entangled, is that the partial 
transpose $\rho_{\mathrm{AB}}^\Gamma$ of their collective 
state $\rho_{\mathrm{AB}}$, has negative eigenvalues. 
A resulting measure of the entanglement 
between A and B is the negativity ${\cal N}$, which 
is the absolute sum of the negative eigenvalues of 
$\rho_{\mathrm{AB}}^\Gamma$. It vanishes for separable 
states, and is equal to $1/2$ for two-qubit maximally 
entangled states \cite{HHHH}. 
To study the entanglement of A and B for low laser 
intensities, i.e. $\Omega \ll \Gamma$, 
we determine the first terms of the expansions, 
in powers of $\eta$, of the eigenvalues of 
$\rho_{\mathrm{AB}}^\Gamma$.

\subsection{Laser intensity threshold for entanglement}

We show here that, as a consequence of the above 
obtained results, A and B are entangled for low enough 
laser intensities. Using expression \eqref{gasstate}, 
the eigenvalues of $\rho_{\mathrm{AB}}^\Gamma$ 
can be evaluated up to second order in $\eta$. 
One of them is close to $1$, and the others, 
denoted $\lambda_q$ in the following, are small, 
since eq.\eqref{gasstate} with $\eta=0$, gives 
$\rho_{\mathrm{AB}}^{(0)}=
| {\cal G} \rangle_{\mathrm{AB}} 
{_{\mathrm{AB}}\langle} {\cal G} |$. Writing  
$\rho_\mathrm{AB}^\Gamma | \varphi_q \rangle 
= \lambda_q | \varphi_q \rangle$, and expanding 
$\rho_\mathrm{AB}^\Gamma$, $| \varphi_q \rangle$ 
and $\lambda_q$, in powers of $\eta$, 
with $\lambda_q^{(0)}=0$, lead to $\lambda_q^{(1)}=0$, 
and $(V+V^\dag)| \varphi_q \rangle^{(0)}=\lambda_q^{(2)}
| \varphi_q \rangle^{(0)}$, where the operator $V$ is given by 
\begin{equation}
V=\sum_{\mu \le n_\mathrm{A} 
< \nu \le n_\mathrm{A}+n_\mathrm{B}} v_{\mu \nu} 
| \mu \rangle_{\mathrm{AB}} 
{_{\mathrm{AB}}\langle} \nu | , \label{V}
\end{equation}
with $n_\mathrm{A}$ and $n_\mathrm{B}$ the numbers 
of atoms, suitably numbered, of systems A and B, 
respectively, and $| \mu \rangle_{\mathrm{AB}} 
= \sigma^\dag_\mu 
\otimes_{\nu \le n_\mathrm{A}+n_\mathrm{B}} |g\rangle_\nu$. 
The complete expression of the second-order 
matrix $(\rho_{\mathrm{AB}}^\Gamma)^{[2]}$ can be found 
in Appendix~\ref{app:Sopttas}.

The eigenvalues of the Hermitian operator $V+V^{\dag}$ are 
real. Since it is traceless, some of them are negative as soon 
as $V \ne 0$. More precisely, the non-zero eigenvalues of 
$V+V^{\dag}$ are $\pm \Lambda_q^{1/2}$ where 
$\Lambda_q$ are the non-zero eigenvalues of both positive 
operators $VV^{\dag}$ and $V^{\dag}V$. 
Consequently, A and B are 
either uncorrelated or entangled. This is similar to the pure 
state case, and results from the fact that, up to second order, 
$\rho_{\mathrm{AB}}$ coincides with a pure state, as 
discussed above. In other words, any two subgroups 
of atoms, are generically entangled for small 
enough $\eta$. The opposite limit, 
$\Omega \gg \Gamma$, corresponds to the saturation 
regime, where $\rho_\mathrm{AB}$ is proportional to 
the identity matrix, and hence A and B are uncorrelated. 
Thus, there is a laser intensity threshold, that depends 
on A and B, where $\rho_\mathrm{AB}$ goes from 
entangled to separable.

\subsection{Dilute regime}

In the general case, determining the value of $\eta$ 
above which $\rho_\mathrm{AB}$ becomes separable, 
requires solving equations \eqref{eqsp} for finite $\eta$, 
which is not straightforward, even for only two atoms 
\cite{ATF}. Moreover, for more than two atoms, there is no 
simple necessary and sufficient condition for entanglement 
\cite{HHHH}. However, for atoms separated by distances 
much larger than $k_0^{-1}$, which is of the order of 
0.1$\mu$m for $\omega_0$ of some eV, 
a laser intensity below which A and B are certainly 
entangled, can be evaluated. In this dilute regime, 
equations \eqref{eqcoeff} can be solved perturbatively 
in the coefficients 
$z_{\mu \nu} \sim k_0^{-1} |{\bf r}_\mu-{\bf r}_\nu|^{-1}$ 
with $\mu \ne \nu$. This leads to the dominant contribution
\begin{equation}
v_{\mu \nu}=-4 z_{\mu \nu} (1-2i\delta)^{-3} 
(w_\mu^2+w_\nu^2 ) , \label{dgr}
\end{equation}
to the matrix elements of operator \eqref{V}. Note that 
the correlations between atoms $\mu$ and $\nu$ are 
then the same in the presence and absence of the other 
atoms. We also remark that atoms not illuminated 
by the laser beam, are also entangled for sufficiently low 
laser intensities. For such atoms, in the dilute regime, 
$v_{\mu \nu} \propto 
\sum_\xi z_{\mu \xi} z_{\nu \xi} w_\xi^2$, 
where the sum runs over the atoms in the laser field.

The eigenvalue $\lambda_q$ expands, in powers 
of $\eta$, as $\lambda_q=\eta^2 \lambda_q^{(2)}
+ \eta^4 \lambda_q^{(4)}+ \ldots$, where 
$\lambda_q^{(2)}$ is an eigenvalue of $V+V^\dag$. 
Since $\lambda_q$ is positive for large $\eta$, 
it changes sign for a certain value $\Omega_q$ 
of $\Omega$, for negative $\lambda_q^{(2)}$. 
As the matrix elements of $V$ are 
given by eq.\eqref{dgr}, $\lambda_q^{(2)}$ is small in 
the dilute regime considered here. 
On the contrary, $\lambda_q^{(4)}$ attains a finite value 
in the limit of vanishing $z_{\mu \nu}$. We find 
the positive asymptotic value
\begin{equation}
\lambda_q^{(4)}=(1/4+\delta^2)^{-2} 
\sum_{\mu \le n_\mathrm{AB}} |w_\mu|^4 
\big|{_\mathrm{AB }\langle} \mu | \varphi_q \rangle^{(0)} \big|^2 , 
\label{lambda4}
\end{equation}
where $n_\mathrm{AB}=n_\mathrm{A}+n_\mathrm{B}$, 
and $| \varphi_q \rangle^{(0)}$ is the eigenstate of $V+V^\dag$ 
corresponding to $\lambda_q^{(2)}$, 
see Appendix~\ref{app:Eelfodr}. 
This leads, for negative $\lambda_q^{(2)}$, to $\Omega_q 
\simeq \Gamma [|\lambda_q^{(2)}|/\lambda_q^{(4)}]^{1/2}$. 
As long as $\Omega<\mathrm{max}_q \Omega_q$, at least 
one eigenvalue $\lambda_q$ is negative, and hence A and B 
are necessarily entangled.

\subsection{Long-range entanglement}\label{sec:Lre}

To study more quantitatively long-range entanglement, 
we consider two regions of characteristic size $L$, 
separated by a large distance $D \gg k_0 L^2$, 
and assume that, in these areas, the laser beam is 
essentially a plane wave of wave vector ${\bf K}$. 
Systems A and B consist of the atoms lying in these regions, 
see Fig.\ref{fig:at}. In this case, equations 
\eqref{gamome}, \eqref{V}, and \eqref{dgr}, give
\begin{equation}
V= \frac{3i\sin^2 \theta e^{i k_0 D}}{k_0 D(1-2i\delta)^3}
\sum_{\mu \le n_\mathrm{A} 
< \nu \le n_\mathrm{AB}} | \tilde \mu \rangle \langle \tilde \nu | 
( w_\mu^2 + w_\nu^2 ) ,  \label{Vds}
\end{equation}  
where $w_\mu=\exp(i{\bf K}\cdot{\bf r}_\mu-i\omega t)$, 
$\theta$ is the angle between ${\bf d}$ and the approximate 
line joining A and B, and $| \tilde \mu \rangle 
= \exp(-i k_0 {\bf e} \cdot {\bf r}_\mu ) | \mu \rangle_{AB}$ 
with ${\bf e}$ the unit vector pointing from A to B. Noting that 
this operator can be written in terms of four kets, one finds 
two negative eigenvalues $\lambda_q^{(2)}$. 
For randomly distributed atoms and large 
enough numbers $n_\mathrm{A}$ and $n_\mathrm{B}$, 
these two negative $\lambda_q^{(2)}$ are practically equal, 
see Appendix~\ref{app:Eoge16}. 
Since $|w_\mu|=1$ for all the atoms of A and B, the sum 
in expression \eqref{lambda4} reduces to $1$. Finally, 
using the evaluation of $\Omega_q$ discussed at the end 
of the previous paragraph, one finds that A and B are 
necessarily entangled for
\begin{equation}
\Omega < \frac{\sqrt{3}}{2}\Gamma (1+4\delta^2)^{1/4} 
\left( \frac{D_0}{D} \right)^{1/2} , \label{Omegamax} 
\end{equation}
where $D_0=k_0^{-1}(n_\mathrm{A} n_\mathrm{B})^{1/2} 
\sin^2 \theta$. The negativity ${\cal N}$ vanishes 
for $\Omega$ equal to the right side of this inequality, 
and also as $\eta$ goes to zero. In the dilute 
regime considered here, it reaches a maximum 
for $\Omega$ equal to the right side 
of eq.\eqref{Omegamax} divided by $\sqrt{2}$, which is 
\begin{eqnarray}
{\cal N}_{max} &=& \frac{9}{32} (1+4\delta^2)^{-1} 
\left( \frac{D_0}{D} \right)^{2} \label{Nmax} \\ 
&=& 32 (1+4\delta^2)^{-2} \eta_{max}^4 \nonumber ,
\end{eqnarray}
where $\eta_{max}$ is the value of the ratio 
$\eta=\Omega/2\Gamma$ at the maximum. 
The attainable values of negativity are thus essentially 
limited by the validity of the low-laser-intensity perturbative 
approach we use. 

As the distance between systems A and B increases, 
the interval of laser amplitudes that lead to non-zero 
negativity, shrinks, and the maximum negativity 
${\cal N}_{max}$ diminishes. However, since $D$ appears 
in equations \eqref{Omegamax} and \eqref{Nmax}, 
divided by $D_0$, the unfavorable impact of increasing 
the distance can be counterbalanced by enlarging 
the numbers $n_\mathrm{A}$ and $n_\mathrm{B}$. 
An interesting consequence is that, all the other 
parameters, including $D$ and the laser amplitude 
$\Omega$, being fixed, big enough groups of atoms 
are necessarily entangled. Let us examine this point 
more carefully. Assuming that the atoms are uniformly 
distributed, and that A and B are cubes of 
edge length $L$, $n_\mathrm{A/B}=(L/d)^3$ 
where $d$ is the mean inter-particle distance, 
and eq.\eqref{Omegamax} can be rewritten as 
\begin{equation}
L > d (1+4\delta^2)^{-1/6} (k_0 D)^{1/3} 
(2\Omega/\sqrt{3}\Gamma \sin \theta)^{2/3} . 
\label{Lmin}
\end{equation}
For $L$ satisfying this inequality, A and B are 
certainly entangled, whereas the negativity vanishes 
for smaller groups of atoms. Note that the above results 
have been obtained under the condition $D \gg k_0 L^2$, 
which can be fulfilled together with eq.\eqref{Lmin} 
only if $D$ is large enough. 
With $k_0^{-1} \simeq 0.1 \mu$m, $d \simeq 1 \mu$m, 
$D \simeq 1$ m (1 cm), $\theta \simeq \pi/2$, 
$\delta \simeq 0$, and $\Omega/\Gamma \simeq 0.1$, 
eq.\eqref{Lmin} gives a lower bound of about 
$50 \mu$m ($10 \mu$m). The corresponding number 
$n_\mathrm{A/B}$ of atoms is of the order 
of $10^5$ ($10^3$). We finally discuss the influence 
of the laser detuning. As $\delta$ is increased, 
the bound given by eq.\eqref{Omegamax} grows, 
and that given by eq.\eqref{Lmin} decreases. However, 
our resonant approach, based on eq.\eqref{eqsp}, is 
valid only for not too large $\delta$. Moreover, 
the reachable values of negativity vanish with 
increasing $\delta$, see eq.\eqref{Nmax}.

\section{Conclusion}

In summary, we have shown that two groups of two-level 
atoms, A and B, can be entangled by a weak resonant laser 
field, even if the distance between them is macroscopic, 
and even in the presence of surrounding identical atoms. 
In a dilute regime, for far separated A and B in a plane 
wave laser beam, we have determined a value of 
the laser amplitude below which A and B are certainly 
entangled, and the maximum negativity that can be 
reached by varying the laser amplitude. They both 
diminish with increasing distance between A and B. 
But these tendencies can be counterbalanced 
by enlarging the sizes of A and B. Consequently, 
for given laser intensity and distance between the two 
groups of atoms, they are necessarily entangled if 
their size exceeds a certain value. 

In this work, we assumed that the motion of the atoms is slow 
enough that its impact on the dynamics of their internal state 
can be disregarded, which, depending on the atomic mass, 
can be valid for temperatures of the order of 10 K. A natural 
extension of our study would be to examine how the found 
laser-induced entanglement depends on the atoms velocities 
for higher temperatures, and whether it disappears at some 
temperature. The quantitative results presented for very 
distant groups of atoms have been derived in the dilute 
regime. It would be of interest to determine how general 
they are, especially the positive impact of enlarging 
the number of considered atoms. We finally remark that, 
though we focus on atoms in this paper, the studied 
entanglement mechanism may be relevant to other 
physical realizations of qubits, such as nuclear spins, 
coupled to a common environment, and to oscillating 
fields. 

\begin{appendix}

\section{Perturbative solution of equation \eqref{eqsp}} 
\label{app:Pse7}

With 
$\rho_p^{(0)}=\delta_{p0} | {\cal G} \rangle \langle {\cal G} |$, 
the recursive relation \eqref{rr} gives 
\begin{equation}
\rho_p^{(1)}=\delta_{p1} \sum_\mu {\tilde u}_\mu 
| \mu \rangle \langle {\cal G} | + \delta_{p \, -1} 
\sum_\mu {\tilde u}^*_\mu | {\cal G} \rangle \langle \mu | ,
\label{rhop1}
\end{equation}
where the components ${\tilde u}_\mu$ obey 
eq.\eqref{eqcoeff} with $t=0$. In deriving this result, we used 
the fact that the matrix elements 
$|\langle \mu \nu | \rho_{\pm 1} | \xi \rangle|<
(\langle \mu \nu | \rho_0 | \mu \nu \rangle 
\langle \xi | \rho_0 | \xi \rangle)^{1/2}$ are at least of second 
order. They are actually of third order. 

Using again eq.\eqref{rr}, we find
\begin{eqnarray}
\rho_p^{(2)}&=&\delta_{p0} \Big( -\sum_\mu |u_\mu|^2 
| {\cal G} \rangle \langle {\cal G} | 
+ \sum_{\mu,\nu} {\tilde u}_\mu{\tilde u}_\nu^* 
| \mu \rangle \langle \nu | \Big) \label{rhop2} \\ &~&+
\delta_{p2} \sum_{\mu<\nu} s_{\mu\nu} 
| \mu \nu \rangle \langle {\cal G} | + \delta_{p \, -2} 
\sum_{\mu<\nu} s^*_{\mu\nu} 
| {\cal G} \rangle \langle \mu \nu | , \nonumber
\end{eqnarray}
where the components $s_{\mu \nu}$ obey 
$\sum_\xi \big( z_{\mu \xi} {\tilde s}_{\xi \nu} 
+ z_{\nu \xi} {\tilde s}_{\xi \mu} \big) -2i\delta s_{\mu \nu} 
= i{\tilde u}_\mu w_\nu + i{\tilde u}_\nu w_\mu$ 
with ${\tilde s}_{\mu \nu}$ equal to $0$ for $\mu=\nu$, 
$s_{\mu \nu}$ for $\mu<\nu$, and $s_{\nu \mu}$ 
for $\mu>\nu$. With the help of the first equality 
of eq.\eqref{eqcoeff}, it can be shown that 
$v_{\mu \nu}=(s_{\mu \nu}-{\tilde u}_\mu{\tilde u}_\nu)
\exp(-2i\omega t)$ satisfies the second equality of 
eq.\eqref{eqcoeff}. Finally, the atoms state 
$\rho=\sum_p{\rho}_p \exp(-ip \omega t)$, can be written, 
up to second order, under the form \eqref{gasstate}.
 
 \section{Expression of $(\rho_\mathrm{AB}^\Gamma)^{[2]}$} 
 \label{app:Sopttas}
 
The Fourier components of the second-order state 
$\rho_\mathrm{AB}^{[2]}$ of two atomic subensembles 
A and B, are readily obtained from eq.\eqref{rhop1} and 
eq.\eqref{rhop2} by performing a partial trace. They are 
given by eq.\eqref{rhop1} and eq.\eqref{rhop2} with sums 
running only over the atoms 
$\mu \le n_\mathrm{AB}=n_\mathrm{A}+n_\mathrm{B}$. 
Consequently, the only non-vanishing matrix elements of 
its partial transpose are 
\begin{eqnarray}
\langle {\cal G} | (\rho_\mathrm{AB}^\Gamma)^{[2]} 
| {\cal G} \rangle
&=& 1-\eta^2 \sum_{\mu \le n_\mathrm{AB}} |u_\mu|^2 , \\
\langle {\cal G} | (\rho_\mathrm{AB}^\Gamma)^{[2]} 
| \mu \rangle
&=& \eta u_\mu^* \;\; \mathrm{for} \;\; 
\mu \le n_\mathrm{A} \nonumber \\
&=& \eta u_\mu  \;\; \mathrm{for} \;\;  
n_\mathrm{A} < \mu , \\
\langle \mu | (\rho_\mathrm{AB}^\Gamma)^{[2]} 
| \nu \rangle &=&
\eta^2 u_\mu u_\nu^*  \;\; \mathrm{for} 
\;\; \mu,\nu \le n_\mathrm{A} \nonumber  \\
&=& \eta^2 u_\mu^* u_\nu  \;\; \mathrm{for} \;\; 
n_\mathrm{A} < \mu,\nu  \\
&=& \eta^2 (u_\mu u_\nu+v_{\mu \nu})  \;\; \mathrm{for} \;\; 
\mu \le n_\mathrm{A} < \nu \nonumber \\
&=& \eta^2 (u_\mu^* u_\nu^*+v_{\nu \mu}^*)  \;\; 
\mathrm{for} \;\; 
\nu \le n_\mathrm{A} < \mu \nonumber , \\
\langle {\cal G} | (\rho_\mathrm{AB}^\Gamma)^{[2]} 
| \mu \nu \rangle 
&=& \eta^2 (u_\mu^* u_\nu^*+v_{\mu \nu}^*) 
\;\; \mathrm{for} \;\; \mu < \nu \le n_\mathrm{A} \nonumber \\
&=& \eta^2 (u_\mu u_\nu + v_{\mu \nu}) \;\; \mathrm{for} \;\; 
n_\mathrm{A} < \mu < \nu  \nonumber \\
&=& \eta^2 u_\mu^* u_\nu  \;\; \mathrm{for} \;\; 
\mu \le n_\mathrm{A} < \nu  ,
\end{eqnarray}
$\langle \mu | (\rho_\mathrm{AB}^\Gamma)^{[2]} 
| {\cal G} \rangle=\langle {\cal G} | 
(\rho_\mathrm{AB}^\Gamma)^{[2]} | \mu \rangle^*$, 
and $\langle \mu \nu | (\rho_\mathrm{AB}^\Gamma)^{[2]} 
| {\cal G} \rangle=\langle {\cal G} | 
(\rho_\mathrm{AB}^\Gamma)^{[2]} | \mu \nu \rangle^*$, 
where $| {\cal G} \rangle$, $| \mu \rangle$ and 
$| \mu \nu \rangle$ must be understood here as 
$| {\cal G} \rangle
=\otimes_{\xi \le n_\mathrm{AB}} | g \rangle_\xi$, 
$| \mu \rangle=\sigma^\dag_\mu | {\cal G} \rangle$ 
and $| \mu \nu \rangle=\sigma^\dag_\mu \sigma^\dag_\nu 
| {\cal G} \rangle$, with $\mu<\nu \le n_\mathrm{AB}$. 

\section{Evaluation of $\lambda_q^{(4)}$ in the dilute 
regime} \label{app:Eelfodr}

Writing  
$\rho_\mathrm{AB}^\Gamma | \varphi_q \rangle 
= \lambda_q | \varphi_q \rangle$, and expanding 
$\rho_\mathrm{AB}^\Gamma$, $| \varphi_q \rangle$ 
and $\lambda_q$, in powers of $\eta$, 
with $\lambda_q^{(0)}=0$, give 
$\lambda_q^{(1)}=\lambda_q^{(3)}{=0}$, 
$(V+V^\dag)| \varphi_q \rangle^{(0)}=\lambda_q^{(2)}
| \varphi_q \rangle^{(0)}$, where the operator $V$ is given 
by eq.\eqref{V}, and, after a lengthy but straightforward 
derivation, 
\begin{eqnarray}
\lambda_q^{(4)} &=& 
\lambda_q^{(2)} \sum_{\mu<\nu} | \tau^{\mu \nu}_q |^2
- |\tau_q^{\cal G}|^2\big( \lambda_q^{(2)} 
+ \sum_\mu |u_\mu|^2 \big) \label{lambdaq4} \\&~& 
+\langle \phi_q |
(\rho_\mathrm{AB}^\Gamma)^{(4)} | \phi_q \rangle
+ 2 \mathrm{Re} \big(\tau_q^{\cal G} \langle \phi_q |
(\rho_\mathrm{AB}^\Gamma)^{(3)} | {\cal G} \rangle \big) 
\nonumber ,
\end{eqnarray}
where $| \phi_q \rangle =| \varphi_q \rangle^{(0)} $, 
$\tau^k_q=\langle k | \varphi_q \rangle^{(1)}$, and the sums 
run only over the atoms $\mu \le n_\mathrm{AB}$. 
In this Appendix, as in the previous one, $| {\cal G} \rangle$, 
$| \mu \rangle$ and $| \mu \nu \rangle$ must be understood 
as $| {\cal G} \rangle
=\otimes_{\xi \le n_\mathrm{AB}} | g \rangle_\xi$, 
$| \mu \rangle=\sigma^\dag_\mu | {\cal G} \rangle$ 
and $| \mu \nu \rangle=\sigma^\dag_\mu \sigma^\dag_\nu 
| {\cal G} \rangle$, with $\mu<\nu \le n_\mathrm{AB}$. 
The only component of $| \varphi_q \rangle^{(1)}$ required 
for our purpose, is $\tau_q^{\cal G}
=-\sum_\mu {\hat u}_\mu \langle \mu |\phi_q \rangle$, 
where ${\hat u}_\mu=u_\mu^*$ for $\mu \le n_\mathrm{A}$, 
and $u_\mu$ for $\mu > n_\mathrm{A}$. 

We are concerned 
with the value of $\lambda_q^{(4)}$ in the limit of infinitely 
distant atoms, in which $\rho_\mathrm{AB}$ 
converges to the uncorrelated state 
$\rho^\mathrm{(dl)}_\mathrm{AB}
=\otimes_{\mu \le n_\mathrm{AB}} \rho_\mu^\mathrm{(dl)}$ 
where
\begin{equation}
\rho_\mu^\mathrm{(dl)}=(1-p_\mu)
\sigma^{\phantom{\dag}}_\mu \sigma_\mu^\dag 
+p_\mu \sigma^\dag_\mu \sigma^{\phantom{\dag}}_\mu
+c_\mu \sigma_\mu^\dag 
+ c_\mu^* \sigma^{\phantom{\dag}}_\mu ,
\end{equation}
with 
$p_\mu=\eta^2|w_\mu|^2/(1/4+\delta^2+2\eta^2|w_\mu|^2)$, 
and 
$c_\mu=(i/2-\delta)p_\mu/\eta w_\mu^*$. Expanding 
this density matrix in $\eta$, gives 
$(\rho_\mathrm{AB}^\Gamma)^{(3)}$ and 
$(\rho_\mathrm{AB}^\Gamma)^{(4)}$ in the infinitely dilute 
regime. Using the resulting expressions, the vanishing 
of $\lambda_q^{(2)}$ in this asymptotic regime, and 
equality \eqref{lambdaq4}, leads to eq.\eqref{lambda4}.

\section{Diagonalisation of the operator $V+V^{\dag}$ 
for $V$ given by equation \eqref{Vds}} 
\label{app:Eoge16}

The operator \eqref{Vds} can be written in the form
\begin{equation}
V=| \phi_{+} \rangle \langle \phi'_{-} | 
+ | \phi'_{+} \rangle \langle \phi_{-} | .
\end{equation} 
The components of the above kets are 
$\langle {\tilde \mu}  | \phi_{\pm} \rangle=(1 \pm \zeta_\mu)/2$, 
and $\langle {\tilde \mu} | \phi'_{\pm} \rangle
=x^\pm_\mu (1 \pm \zeta_\mu)/2$, where $\zeta_\mu=1$ 
for $\mu \le n_\mathrm{A}$, and $-1$ for 
$\mu > n_\mathrm{A}$, $x^+_\mu
=x \exp( i k_0 D+2i{\bf K}\cdot{\bf r}_\mu-2i\omega t)$ 
with $x=3i\sin^2 \theta/k_0 D(1- 2i\delta)^3$, and 
$x_\mu^-=(x^+_\mu)^*$. To find the non-zero $\lambda$ 
obeying the eigenvalue equation 
$(V+V^\dag)|\varphi \rangle=\lambda |\varphi \rangle$, 
we expand $|\varphi \rangle$ on the basis 
$\{ | \phi_{+} \rangle , | \phi'_{+} \rangle , 
| \phi_{-} \rangle , | \phi'_{-} \rangle  \}$. This leads to 
the characteristic equation
\begin{equation}
\lambda^4-2y\big[1+
\mathrm{Re}(s_\mathrm{A}s_\mathrm{B}^*)\big]
\lambda^2+y^2 \big[ 1-|s_\mathrm{A}|^2 \big]
\big[ 1-|s_\mathrm{B}|^2 \big] =0 ,
\end{equation}
where $y=n_\mathrm{A}n_\mathrm{B}|x|^2$, 
$s_\mathrm{A}=n_\mathrm{A}^{-1}
\sum_{\mu \le n_\mathrm{A}} 
\exp(2i{\bf K}\cdot{\bf r}_\mu)$ 
and $s_\mathrm{B}=n_\mathrm{B}^{-1}
\sum_{\mu > n_\mathrm{A}} \exp(2i{\bf K}\cdot{\bf r}_\mu)$. 

For randomly distributed atoms and large enough numbers 
$n_\mathrm{A}$ and $n_\mathrm{B}$, 
$s_\mathrm{A}$ and $s_\mathrm{B}$ are 
negligible, and the above equation 
simplifies to $(\lambda^2-y)^2=0$. This last result can 
be derived more directly as follows. The relations 
$s_\mathrm{A}, s_\mathrm{B} \ll 1$ 
can be rewritten as 
$\langle \phi_{\pm} | \phi'_{\pm} \rangle \simeq 0$. 
When these products vanish, it is immediate to see 
that the non-zero eigenvalues of both 
$VV^{\dag}$ and $V^{\dag}V$, which are the squares 
of the non-zero eigenvalues of $V+V^\dag$, are 
$\langle \phi_\pm  | \phi_{\pm} \rangle 
\langle \phi'_\mp  | \phi'_{\mp} \rangle=y$.
 
\end{appendix}


\begin{thebibliography}{99}

\bibitem{YE} 
T. Yu and J.H. Eberly, Phys. Rev. Lett. {\bf 93}, 140404 (2004)

\bibitem{Br} D. Braun, Phys. Rev. Lett. {\bf 89}, 
277901 (2002).

\bibitem{J} L. Jak\' obczyk, J.Phys. A: Math. Gen. 
{\bf 35}, 6383 (2002). 

\bibitem{BFP} F. Benatti, R. Floreanini and M. Piani, 
Phys. Rev. Lett. {\bf 91}, 070402 (2003).

\bibitem{FT} Z. Ficek and R. Tana{\'s}, 
Phys. Rev. A {\bf 74}, 024304 (2006).

\bibitem{MCNBF} D. P. S. McCutcheon, A. Nazir, 
S. Bose and  A. J. Fisher, Phys. Rev. A {\bf 80}, 
022337 (2009).

\bibitem{CKS} {\" O}. {\c C}akir, A. A. Klyachko and 
A. S. Shumovsky, Phys. Rev. A {\bf 71}, 034303 (2005).

\bibitem{ATF} K. Almutairi, R. Tana\'s and Z. Ficek, 
Phys. Rev. A {\bf 84}, 013831 (2011).

\bibitem{EPJB} 
S. Camalet, Eur. Phys. J. B {\bf 84}, 467 (2011)

\bibitem{BA1} 
B. Bellomo and M. Antezza, epl {\bf 104}, 10006 (2013)

\bibitem{BA2} 
B. Bellomo and M. Antezza, New J. Phys. {\bf 15}, 
113052 (2013)

\bibitem{CDG2} 
C. Cohen-Tannoudji, J. Dupont-Roc and G. Grynberg,
{\it Processus d'interaction entre photons et atomes}, 
(CNRS Editions, Paris, 1988)

\bibitem{D} 
R.H. Dicke, Phys. Rev. {\bf 93}, 99 (1954)

\bibitem{L2} 
R.H. Lehmberg, Phys. Rev. A {\bf 2}, 889 (1970)

\bibitem{Wea} 
T. Wilk et al, Phys. Rev. Lett. {\bf 104}, 010502 (2010)

\bibitem{Zea} 
X.L. Zhang et al, Phys. Rev. A {\bf 82}, 030306(R) (2010)

\bibitem{PSC} 
A.S. Parkins, E. Solano and J.I. Cirac, Phys. Rev. Lett. 
{\bf 96}, 053602 (2006)

\bibitem{MPC} 
C.A. Muschik, E.S. Polzik and J.I. Cirac, Phys. Rev. A {\bf 83}, 
052312 (2011)

\bibitem{RKS} 
F. Reiter, M.J. Kastoryano, and A.S. S\o rensen, 
New J. Phys. {\bf 14}, 053022 (2012)

\bibitem{KM} 
H. Krauter et al, Phys. Rev. Lett. {\bf 107}, 080503 (2011)

\bibitem{MK} 
C.A. Muschik et al, J. Phys. B: At. Mol. Opt. Phys. {\bf 45}, 
124021 (2012)

\bibitem{MEK} M. Macovei, J. Evers, and C.H. Keitel, 
J. Mod. Opt. {\bf 57}, 1287 (2010)

\bibitem{EPJB2} 
S. Camalet, Eur. Phys. J. B {\bf 86}, 176 (2013)

\bibitem{L1} 
R.H. Lehmberg, Phys. Rev. A {\bf 2}, 883 (1970)

\bibitem{DC} J. Dalibard and C. Cohen-Tannoudji, 
J. Opt. Soc. Am. B {\bf 2}, 1707 (1985) 

\bibitem{S} 
M. J. Stephen, J. Chem. Phys. {\bf 40}, 669 (1964)

\bibitem{TGCB} 
M. Trippenbach, B. Gao, J. Cooper and K. Burnett, 
Phys. Rev. A {\bf 45}, 6539 (1992)

\bibitem{G} C.C. Gerry, Phys. Rev. A {\bf 53}, 
4583 (1996)

\bibitem{vE} S.J. van Enk, Phys. Rev. A {\bf 72}, 
064306 (2005)

\bibitem{CDLK} K.S. Choi, H. Deng, J. Laurat, and 
 H.J. Kimble, Nature {\bf 452}, 67 (2008) 

\bibitem{HHHH} R. Horodecki, P. Horodecki, M. Horodecki 
and K. Horodecki, Rev. Mod. Phys. {\bf 81}, 865 (2009)

\end{thebibliography}
\end{document}